\documentclass[12pt]{article}
\usepackage{epsf}
\def\beq{\begin{equation}}
\def\eeq{\end{equation}}
\def\ap#1#2#3 {Ann. Phys. (NY) {\bf#1} (19#2) #3}
\def\err#1#2#3 {{\it Erratum} {\bf#1} (19#2) #3}
\def\ib#1#2#3 {{\it ibid.} {\bf#1} (19#2) #3}
\def\ijmp#1#2#3 {Int. J. Mod. Phys. {\bf#1} (19#2) #3}
\def\jetp#1#2#3 {JETP Lett. {\bf#1} (19#2) #3}
\def\mpl#1#2#3 {Mod. Phys. Lett. {\bf#1} (19#2) #3}
\def\np#1#2#3 {Nucl. Phys. {\bf#1} (19#2) #3}
\def\pl#1#2#3 {Phys. Lett. {\bf#1} (19#2) #3}
\def\prep#1#2#3 {Phys. Rep. {\bf#1} (19#2) #3}
\def\prev#1#2#3 {Phys. Rev. {\bf#1} (19#2) #3}
\def\prl#1#2#3 {Phys. Rev. Lett. {\bf#1} (19#2) #3}
\def\sjnp#1#2#3 {Sov. J. Nucl. Phys. {\bf#1} (19#2) #3}
\def\spj#1#2#3 {Sov. Phys. JETP {\bf#1} (19#2) #3}
\def\spu#1#2#3 {Sov. Phys. Usp. {\bf#1} (19#2) #3}
\def\zp#1#2#3 {Zeit. Phys. {\bf#1} (19#2) #3}
\def\a{\alpha}

\begin{document}
\begin{titlepage}
\begin{center}
{\Large \bf William I. Fine Theoretical Physics Institute \\
University of Minnesota \\}  \end{center}
\vspace{0.2in}
\begin{flushright}
TPI-MINN-03/33-T \\
UMN-TH-2221-03 \\
November 2003 \\
\end{flushright}
\vspace{0.3in}
\begin{center}
{\Large \bf  The photon spectrum in orthopositronium decay at
$\omega_\gamma \ll m_e$
\\}
\vspace{0.2in}
{\bf M.B. Voloshin  \\ }
William I. Fine Theoretical Physics Institute, University of
Minnesota,\\ Minneapolis, MN 55455 \\
and \\
Institute of Theoretical and Experimental Physics, Moscow, 117259
\\[0.2in]
\end{center}

\begin{abstract}
A closed analytic expression is given for the spectrum of low energy
photons in the annihilation of orthopositronium, which expression sums
all the effects of the Coulomb interaction between the electron and the
positron. The applicability of the formula is limited only by the
condition $\omega_\gamma \ll m_e$. In the region $\omega_\gamma \gg m_e
\alpha^2$ the Coulomb interaction term gives the leading at low energy
one-loop correction, proportional to $\alpha \sqrt{m_e/\omega_\gamma}$,
to the decay spectrum. The constant in $\omega_\gamma$ radiative term in
the one-loop correction to the spectrum is also presented here in an
analytic form.
\end{abstract}

\end{titlepage}

\section{Introduction}
The interplay of binding and radiative effects in the properties of
positronium makes it an interesting case study in pure QED.  In
particular, the annihilation of the lowest $^3S_1$ state of
orthopositronium (o-Ps) into three photons has been studied both
experimentally and theoretically since the first calculation of this
process by Ore and Powell\cite{op}. Most recently Manohar and
Ruiz-Femen\'\i a\cite{mrf} have considered the effects of the Coulomb
interaction on the photon spectrum in the o-Ps decay in the region of
small photon energies, where these effects are essential. They have
found in an integral form the expression for this spectrum, which takes
into account to all orders the Coulomb interaction between the electron
and the positron in the nonrelativistic approximation, and they also
argued that their result is applicable in the region of the photon
energy, $\omega$, satisfying the condition $\omega \ll m \alpha$, with
$m$ being the mass of the electron. The present paper contains
essentially two further developments over the results of Ref.\cite{mrf}:
a closed analytic formula in terms of just one standard hypergeometric
function is given for the spectrum of low energy photons, and it is
shown that the condition for its applicability is in fact $\omega \ll
m$, i.e. the formula is valid in a significantly broader range of
energies than stated in Ref.\cite{mrf}. Furthermore, in the region
$\omega \gg  m \alpha^2$, the expansion parameter for the Coulomb
effects is $\alpha \sqrt{m/\omega}$, and the first term of this
expansion describes the low-energy asymptotic behavior of the one-loop
QED corrections to the spectrum. Due to the rather slow rise of the
Coulomb term at low energy it is interesting to also evaluate the
constant in $\omega$ term at small $\omega$ in the one-loop correction.
As will be shown here this term can in fact be found from the known in
the literature radiative QED corrections to the total rates of decay of
the $C$ even $^1S_0$ and $^3P_{0,2}$ states of positronium into two
photons.  A comparison of the resulting expression for the first two
terms of the expansion in $\omega/m$ of the $O(\a)$ correction to the
spectrum in the orthopositronium decay with the only available results
of a numerical calculation\cite{adkins} of this correction is also
discussed here.

\section{Multipole expansion at $\omega \ll m$}
The spectrum of low-energy photons at $\omega \ll m$ lends itself to a
nonrelativistic treatment\footnote{A similar treatment in QCD was
used\cite{mv1} in discussion of soft gluons in a three-gluon
annihilation of heavy quarkonium.}. Indeed, after emission of such soft
photon from the initial $^3S_1$ state of o-Ps at energy $E_0=-m \a^2/4$
(relative to the threshold), the (virtual) $e^+e^-$ pair remains
nonrelativistic and is in a $C$-even state and has negative energy
$E=E_0-\omega$, which state then  annihilates into two hard photons.
This picture can be represented\cite{mv1,mrf} by the diagram (of the
nonrelativistic perturbation theory) shown in Fig.1. The annihilation
into two hard gluons takes place at distances $O(m^{-1})$, which are
considered as infinitesimal  in terms of the nonrelativistic relative
motion of the electron and the positron. The Green's function at energy
$E$, describing the propagation of the pair between the emission of the
soft photon and annihilation, thus contains the exponential factor
$\exp(-\kappa r)$ with $\kappa$ defined as $-\kappa^2/m = E$, so that
the discussed process is determined by distances of order $\kappa^{-1}$
and the typical velocities of the electron and positron are given by $v
\sim \kappa/m$. It is well known that in such situation the expansion
parameter for the Coulomb interaction is $\a m/\kappa$. Thus in the
region where $\kappa$ is comparable with $m \a$ one has to use the exact
Green's function in the Coulomb potential. In the region of larger
$\kappa$: $\kappa \gg m \a$, the nonrelativistic treatment is still
applicable as long as $\kappa^2 \ll m^2$, and the Coulomb effects can be
calculated either by a perturbative expansion of the Green's function,
or by an expansion of the exact formula, if such formula is available.

\begin{figure}[ht]
  \begin{center}
    \leavevmode
    \epsfbox{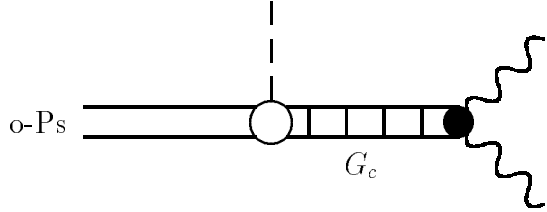}
    \caption{The diagram for the description of the photon spectrum in
the region $\omega \ll m$ in the decay of o-Ps.  The open circle stands
for the interaction of a nonrelativistic $e^+e^-$ pair with a soft
photon (dashed line), and the filled circle shows the annihilation into
two hard photons (wavy lines) at distances $O(m^{-1})$. $G_c$ stands for
the Green's function in the Coulomb field.}
  \end{center}
\label{fig:xy}
\end{figure}

The Hamiltonian for the emission of the soft photon in the diagram of
Fig.1 can be expanded in multipoles. The dominance of the lowest
multipoles is guaranteed by a general consideration\cite{ll} for as long
as the system is nonrelativistic, i.e. $\omega \ll m$. However, since at
this point the present paper differs from Ref.\cite{mrf}, it is
appropriate to provide here a more detailed discussion. The $C$-even
state in Fig.1 annihilating into two hard gluons can be either a
spin-triplet state with an odd orbital momentum $L=2n+1$, or a
spin-singlet state with an even orbital momentum $L=2n$, where in both
cases $n$ is a non-negative integer. For the first set of states the
minimal multipole contributing to the transition $^3S \to ^3(2n+1)
+\gamma$ is of the electric type: $E(2n+1)$, and the amplitude of the
transition is proportional to $(\omega r)^{(2n+1)}$, while for the
latter states the lowest multipole is of the magnetic type: $M(2n)$, and
the amplitude of the transition is proportional to $(\omega/m)\, (\omega
r)^{(2n)}$. The amplitude of the annihilation of a state with orbital
momentum $L$ at distances $O(m^{-1})$ contains $L$ derivatives of the
wave function at the origin, i.e. it is proportional to $(\kappa/m)^L$.
Multiplying the indicated factors for both sets of the intermediate
states, and taking into account that the $\exp(-\kappa r)$ behavior of
the Green's function constrains the product $\kappa r$ at order one, one
readily finds in both cases that the contribution of the corresponding
intermediate state in the amplitude described by Fig.1 contains the
factor $(\omega/m)^{(2n+1)}$. Thus in the nonrelativistic region of
$\omega \ll m$ it is sufficient to consider only the intermediate states
with the lowest $n$, i.e. $n=0$. Clearly, these states are the $^1S$ and
$^3P$, and they are reached from the initial $^3S$ state by respectively
$M1$ and $E1$ radiative transitions. It is important to emphasize that
both these intermediate states provide contribution of the same order in
the nonrelativistic limit\cite{mv1,mrf}, and also to notice that the
spatial extent, $\sim (m \a)^{-1}$, of the initial state of positronium
does not enter as a parameter in the discussed multipole
expansion\footnote{The distances in the radiative transition amplitude
are constrained by the falloff of the Green's function, rather than by
the falloff of the wave function of the initial state. In particular at
$\kappa \gg m \a$ the initial state wave function enters only through
its value at the origin.}

\section{The Coulomb interaction effect in the spectrum}

Using the standard expressions for the Hamiltonian of the $M1$ and $E1$
interaction, and also using the well known amplitudes of the two-photon
annihilation of the $^1S\,$\cite{pomeranchuk} and $^3P\,$\cite{alekseev}
states, one can write the expression for the differential rate of the
o-Ps annihilation in the limit $x \equiv \omega/m \ll 1$ in the
form\cite{mrf}
\beq
{{\rm d} \Gamma \over {\rm d} x} = {m \a^6 \over 9 \pi} x \, \left [
|a_m|^2 + {7 \over 3} |a_e|^2 \right ]~,
\label{amae}
\eeq
where $a_m$ and $a_e$ are respectively the magnetic and electric dipole
amplitudes, which can be written in terms of the wave function and of
the discussed Coulomb Green's function $G_c({\bf x}, {\bf y};
-\kappa^2/m)$ as follows,
\beq
a_m={\omega \over \psi_0(0)} \, \int G_c({\bf 0}, {\bf y}; -\kappa^2/m)
\, \psi_0({\bf y}) \, {\rm d}^3 y~,
\label{am}
\eeq
\beq
a_e={\omega \over 3 \psi_0(0)} \, \int y_i \left . \left [ {\partial
\over \partial x_i} G_c({\bf x}, {\bf y}; -\kappa^2/m) \right ] \right
|_{{\bf x}=0}\, \psi_0({\bf y}) \, {\rm d}^3 y~.
\label{ae}
\eeq
Here $\psi_0({\bf y})$ stands for the wave function of the initial
state of the orthopositronium, and the factor $\psi_0(0)$ appears in the
denominator in both these formulas due to that its value is already
included in the normalization in eq.(\ref{amae}). In other words,  the
amplitudes $a_m$ and $a_e$ are normalized in such a way that they both
are equal to one if the Coulomb Green's function $G_c$ is replaced by
the free motion one $G_f$, in which limit the lowest-order
formula\cite{op} for the spectrum is reproduced.

The magnetic amplitude $a_m$ is however trivial and is equal to one also
if the Coulomb interaction is taken into account. This clearly is a
consequence of that the spatial wave function of the ground $^3S$ state
is orthogonal to those of all the $^1S$ states except for the ground
one, where the overlap integral is equal to one. The relation $a_m=1$ is
valid up to relativistic terms, including the $^3S - ^1S$ hyperfine
splitting, which effects are beyond the intended accuracy\footnote{The
hyperfine splitting, becoming essential at very small $\omega$: $\omega
\sim m \a^4$, was considered in Ref.\cite{mrf} in order to establish the
low-energy behavior mandated by the Low theorem.}.

The integral in eq.(\ref{ae}) can be calculated using the partial wave
expansion of the Green's function
\beq
G_c({\bf x}, {\bf y}; E)=\sum_{\ell} (2 \ell +1)\, G_\ell (x, y; E)
P_\ell \left ( {{\bf x} \cdot {\bf y} \over x \, y } \right )~,
\label{gell}
\eeq
with $P_\ell(z)$ being the Legendre polynomials, and the following
representation\cite{hostler} of the partial wave Green's functions
$G_\ell$ in the Coulomb problem:
\beq
G_\ell \left ( x, y; {-\kappa^2 \over m} \right )={m \, \kappa \over 2
\pi} \, (2 \kappa
x)^\ell \, (2\kappa y)^\ell \, e^{-\kappa (x+y)}\, \sum_{n=0}^\infty
{L_n^{(2\ell+1)}(2 \kappa x)\, L_n^{(2\ell+1)}(2 \kappa y) \, n! \over
(n+l+1-\nu) \, (n+2 \ell +1)!}~,
\label{gcoul}
\eeq
where $\nu = m \a /(2 \kappa)$, and $L_n^{(p)}(z)$ are the Laguerre
polynomials defined as
\beq
L_n^{(p)}(z)={e^z z^{-p} \over n!} \, \left ( {{\rm d} \over {\rm d} z}
\right )^n e^{-z} z^{n+p}~.
\label{lag}
\eeq
(Usage of this representation can also be found in similar problems in
Refs. \cite{mv1} and \cite{mv2}.)

The expression in eq.(\ref{ae}) contains only the $P$ wave partial
Green's function $G_1$ and thus can be written as
\begin{eqnarray}
&&a_e(\omega)={4 \pi \, \omega \over \psi_0(0)} \int_0^\infty
G_1 \left ( 0,y;{-\kappa^2 \over m} \right ) \,  \psi_0(y) \, y^4 \,
{\rm d} y = \nonumber \\
&&{1-\nu^2 \over 24} \, \sum_{n=0}^\infty {1 \over n+2 - \nu} \,
\int_0^\infty \, \exp \left ( - {1+\nu \over 2} \, z \right ) \,
L_n^{(3)}(z) \, z^4 \, {\rm d} z =  \nonumber \\
&&{64 \over (1+\nu)^4} \, \sum_{n=0}^\infty {1 \over n+2 - \nu} \, \left
[ {(n+4)! \over 24 \, n!} \left ( - {1-\nu \over 1+ \nu} \right )^n -
{1+ \nu \over 2} \, {(n+3)! \over 6 \, n!} \, \left ( - {1-\nu \over 1+
\nu} \right )^n \right ]~.
\label{rsum}
\end{eqnarray}
Here a use is made of the explicit form of the ground state wave
function: $\psi_0(y)/\psi_0(0)=\exp(-m \a y/2)$ as well as of the
relation $m \omega/\kappa^2=1-\nu^2$. In the last transition the formula
(\ref{lag}) is used to perform integration by parts. The sum in the
latter expression in eq.(\ref{rsum}) is of the Gauss hypergeometric type
and can be done explicitly in terms of the hypergeometric function
$_2F_1$, so that the final result can be written as
\beq
a_e(\omega)={(1-\nu)(3+5 \nu) \over 3 \, (1+\nu)^2} + {8\, \nu^2 \,
(1-\nu) \over 3 \, (2-\nu) \, (1+ \nu)^3 } \, \, _2F_1 \left ( 2-\nu, 1;
3-\nu; - {1-\nu \over 1+
\nu} \right )~.
\label{res}
\eeq
The explicit relation between the photon energy $\omega$ and the Coulomb
parameter $\nu$ reads as
\beq
\omega= {m \a^2 \over 4} \, {1-\nu^2 \over \nu^2}={1 \over 2}\, {\rm
Ry}\, {1-\nu^2 \over \nu^2}~.
\label{omnu}
\eeq

The amplitude $a_e$ given by the formula (\ref{res}) is fully equivalent
to its integral representation described in Ref.\cite{mrf}.

\section{One-loop QED correction to the photon spectrum at $\omega \ll
m$}
The result in eq.(\ref{res}) can be expanded in $\nu$. The linear
term in this expansion describes the correction of the first order in
$\a$:
\beq a_e=1-{4 \over 3}\, \nu + O(\nu^2)= 1- {2 \a \over 3} \,
\sqrt{m \over \omega}+ O(\a^2)~.
\label{lin}
\eeq
When used in eq.(\ref{amae}) for the differential decay rate the linear
term in this expansion determines the asymptotic behavior of the
one-loop QED correction to the spectrum at $\omega \ll m$. However, the
slow $1/\sqrt{x}$ rise of this correction toward small $x$ makes
potentially important the higher terms of the expansion in $x$ of the
$O(\a)$ one-loop correction. Clearly, the relativistic expansion for the
discussed terms due to the Coulomb interaction goes in powers of
$\kappa^2/m^2 \approx \omega/m =x$, so that the next term of this type
is proportional to $\a \sqrt{x}$. The same behavior is true for the
correction terms arising from the full relativistic scattering kernel
for the initial state and from the process (o-Ps)$\, \to \gamma^* \to 3
\gamma$. Although the latter two mechanisms together account for almost
85\% of the $O(\a)$ correction to the total rate\cite{cls}, in the
spectrum at low $x$ their relative contribution starts only as $\a
\sqrt{x}$.  Such terms are beyond the scope of this paper.

On the other hand, the constant in $x$ term, i.e. of order $\a x^0$ at
small $x$, can be quite readily deduced from the known in the literature
results for the one-loop corrections to the total rates of the two
photon annihilation of the $C$-even $S$ and $P$ states of the
orthopositronium. These corrections are of genuinely radiative nature
and arise from distances of order $1/m$, as opposed to the so far
discussed Coulomb effects determined by the electron-positron
interaction at distances of order $1/\kappa$. The sources of such
correction terms can be easily identified from the graph of Fig.1.
Indeed, the elements of  the considered process determined by the
distances of order $1/m$ are the amplitude of the annihilation into two
hard photons and the electron magnetic moment entering the Hamiltonian
of the $M1$ interaction. Accordingly, in order to include the discussed
radiative corrections, the equation (\ref{amae}) should be rewritten as
\beq
{{\rm d} \Gamma \over {\rm d} x} = {m \a^6 \over 9 \pi} x \, \left \{
|a_m|^2 \, \left ( {g_e \over 2} \right )^2 \, {\Gamma(^1S_0 \to
2\gamma) \over \Gamma_0(^1S_0 \to 2\gamma)} +  |a_e|^2 \, \left
[{\Gamma(^3P_0 \to 2\gamma) \over \Gamma_0(^3P_0 \to 2\gamma)}+ {4 \over
3} \, {\Gamma(^3P_2 \to 2\gamma) \over \Gamma_0(^3P_2 \to 2\gamma)}
\right ] \right \}~,
\label{amae1}
\eeq
where $g_e$ is the electron gyromagnetic ratio, and the ratia of the
decay rates for the indicated $C$-even states are to their values
($\Gamma_0$) in the lowest order\footnote{There is of course no
contribution from the $^3P_1$ state due to the Landau-Yang theorem.}.
The $O(\a)$ terms in these ratia are known in the literature:
\beq
{\Gamma(^1S_0 \to 2\gamma) \over \Gamma_0(^1S_0 \to 2\gamma)}=1+{\a
\over \pi} \, \left ({\pi^2 \over 4} -5 \right )
\label{ps}
\eeq
for the parapositronium decay\cite{hb}, and
\beq
{\Gamma(^3P_0 \to 2\gamma) \over \Gamma_0(^3P_0 \to 2\gamma)}=1+{\a
\over \pi} \, \left ({\pi^2 \over 4} -{7 \over 3} \right ),
~~~{\Gamma(^3P_2 \to 2\gamma) \over \Gamma_0(^3P_2 \to 2\gamma)}=1 - {4
\a \over \pi}
\label{3p}
\eeq
for the $^3P_{0,2}$ states, which formulas are an adaptation to QED of
the quarkonium results for QCD corrections in Ref.\cite{bcgr}.

Using the expressions (\ref{ps}) and (\ref{3p}) and also the famous
Schwinger's result $g_e/2=1+\a/(2\pi) + O(\a^2)$ in eq.(\ref{amae1}),
the formula for the spectrum can be written including the radiative
correction:
\begin{eqnarray}
&&{{\rm d} \Gamma \over {\rm d} x} = {m \a^6 \over 9 \pi} x \, \left \{
|a_m|^2 \, \left [1+{\a \over \pi}\, \left ({\pi^2 \over 4}-4
\right)\right] + |a_e|^2 \, \left[{7 \over 3}+{\a \over \pi}\, \left
({\pi^2 \over 4}-{23 \over 3} \right)\right]  \right \} \nonumber \\
&&={10 \, m \a^6 \over 27 \pi} x \, \left [ 1 - {\a \over \pi}\, \left
({14 \pi \over 15 \, \sqrt{x}} -{3 \pi^2 \over 20} + {7 \over 2} \right
) + O(\a^2) \right]~,
\label{amae2}
\end{eqnarray}
where the last expression also includes the first Coulomb correction to
$a_e$ from eq.(\ref{res}).

A general calculation of the one-loop QED correction to the spectrum in
o-Ps decay has been done numerically by Adkins\cite{adkins}, and it is
instructive to compare the two results. For the purpose of such
comparison, following the conventions of Ref.\cite{adkins}, we write the
formula for the differential probability in the form
\beq
{1 \over \Gamma_0} \, {{\rm d} \Gamma \over {\rm d} x}=\gamma_0(x) + {\a
\over \pi} \, \gamma_1(x)~.
\label{norms}
\eeq
Here $\Gamma_0=2 \, (\pi^2-9) \, m \a^6/(9 \pi)$ is the total decay
rate in the lowest order\cite{op}, $\gamma_0$ is the normalized
differential decay rate in the same lowest order, for which we use here
its nonrelativistic limit at small $x$: $\gamma_0(x)=(5/3) x
/(\pi^2-9)$, and $\gamma_1(x)$ is the first-order QED correction to the
spectrum. From the equation (\ref{amae2}) one readily
finds that under the described conventions $\gamma_1(x)$ is given by
\beq
\gamma_1(x)=-{1 \over 9 \, (\pi^2-9)} \, \left [ 14 \pi \, \sqrt{x}+
15 \, \left( {7 \over 2}-{3 \pi^2 \over 20} \right ) x \right ]~.
\label{g1}
\eeq
In Ref.\cite{adkins} the physical region of $x$ from $x=0$ to $x=1$ is
divided into 20 equal bins of 0.05 each, and the integral of
$\gamma_1(x)$ over each of the bins is tabulated. In particular in the
first two bins the integrals are found as respectively -0.0502(11)
and -0.0994(14). The formula in eq.(\ref{g1}) gives for the same
integrals numerical values of -0.0467 and -0.0911. The difference from
the result of the full calculation\cite{adkins}, albeit numerical,
is in a reasonable agreement with the accuracy
expected from using only the first two terms of the expansion in
$\sqrt{x}$ at $x \approx 0.05 - 0.1$.

\section*{Acknowledgments}
I thank Andrzej Czarnecki for pointing out to me the paper
\cite{adkins}. This work is supported in part
by the DOE grant DE-FG02-94ER40823.

\end{document}